\documentclass[aps,prd,preprint,superscriptaddress,tightenlines,%
   nofootinbib]{revtex4}
\newcommand{\PRE}[1]{{#1}}   

\usepackage{bm}
\usepackage{epsfig}

\newcommand{\postscript}[2]{\setlength{\epsfxsize}{#2\hsize}
   \centerline{\epsfbox{#1}}}

\newcommand{\mplanck}{M_{\text{Pl}}}
\renewcommand{\mp}{M_p}
\newcommand{\mstar}{M_{*}}
\newcommand{\lstar}{L_{*}}
\newcommand{\md}{M_D}

\newcommand{\mpmin}{M_p^{\text{min}}}

\newcommand{\gev}{\text{GeV}}
\newcommand{\tev}{\text{TeV}}

\newcommand{\mb}{\text{mb}}
\newcommand{\cm}{\text{cm}}

\newcommand{\g}{\text{g}}

\newcommand{\etal}{{\em et al.}}
\newcommand{\eg}{{\em e.g.}}
\newcommand{\ie}{{\em i.e.}}

\newcommand{\eqref}[1]{Eq.~(\ref{#1})}

\begin{document}

\preprint{
\hfil
\begin{minipage}[t]{3in}
\begin{flushright}
\vspace*{.4in}
NUB--3226--Th--02\\
MIT--CTP--3245\\
UCI--TR--2002--5\\
hep-ph/0202124
\end{flushright}
\end{minipage}
}

\title{
\PRE{\vspace*{1.5in}}
$\bm{p}$-Branes and the GZK Paradox
\PRE{\vspace*{0.3in}}
}

\author{Luis A.~Anchordoqui}
\affiliation{Department of Physics,\\
Northeastern University, Boston, MA 02115
\PRE{\vspace*{.1in}}
}
\author{Jonathan L.~Feng}
\affiliation{Center for Theoretical Physics,\\
Massachusetts Institute of Technology, Cambridge, MA 02139
\PRE{\vspace*{.1in}}
}
\affiliation{Department of Physics and Astronomy,\\
University of California, Irvine, CA 92697
\PRE{\vspace*{.3in}}
}

\author{Haim Goldberg%
\PRE{\vspace*{.2in}}
}
\affiliation{Department of Physics,\\
Northeastern University, Boston, MA 02115
\PRE{\vspace*{.1in}}
}


\begin{abstract}
\PRE{\vspace*{.1in}} 
In spacetimes with asymmetric extra dimensions, cosmic neutrino
interactions may be extraordinarily enhanced by $p$-brane production.
Brane formation and decay may then initiate showers deep in the
Earth's atmosphere at rates far above the standard model rate.  We
explore the $p$-brane discovery potential of cosmic ray experiments.
The absence of deeply penetrating showers at AGASA already provides
multi-TeV bounds on the fundamental Planck scale that significantly
exceed those obtained from black hole production in symmetric
compactification scenarios.  This sensitivity will be further enhanced
at the Auger Observatory.  We also examine the possibility that
$p$-brane formation resolves the GZK paradox.  For flat
compactifications, astrophysical bounds exclude this explanation.  For
warped scenarios, a solution could be consistent with the absence of
deep showers only for extra dimensions with fine-tuned sizes well
below the fundamental Planck length.  In addition, it requires
moderately penetrating showers, so far not reported, and $\sim 100\%$
modifications to standard model phenomenology at 100 GeV energies.
\end{abstract}


\maketitle

A spectacular prediction of scenarios with strong gravity and large
(or warped) extra dimensions~\cite{Antoniadis:1990ew} is the
production of microscopic black holes (BHs) in particle collisions
with center-of-mass energies larger than a TeV~\cite{Banks:1999gd}.
Cosmic neutrinos with energies above $10^{6}~\gev$ that strike a
nucleon in the Earth's atmosphere may then create BHs with cross
sections two or more orders of magnitude above their standard model
cross sections~\cite{Feng:2001ib}. Criticisms~\cite{Voloshin:2001vs}
of the assumptions leading to these cross sections have been
addressed~\cite{Dimopoulos:2001qe}. These BHs are expected to decay
promptly, initiating spectacular quasi-horizontal air showers deep in
the atmosphere.  The distinctive features of BH evaporation allow BHs
to be differentiated from background~\cite{Anchordoqui:2001ei}, and
the production and subsequent evaporation of such BHs may be studied
in detail at cosmic ray
observatories~\cite{Feng:2001ib,Emparan:2001kf,Anchordoqui:2001cg}.
Additionally, neutrinos that traverse the atmosphere unscathed may
produce BHs through interactions in the Earth; detailed
simulations~\cite{Kowalski:2002gb} of these BH events also find
observable rates at neutrino telescopes.

Recently, based on the absence of a significant signal of deeply
developing showers reported by the AGASA Collaboration~\cite{agasa},
we derived new limits on the fundamental Planck scale in spacetimes
with extra dimensions of equal length~\cite{Anchordoqui:2001cg}. More
recently, it was pointed out that for TeV-scale gravity with
asymmetric large extra dimensions~\cite{Lykken:1999ms} the formation
of $p$-branes could be competitive with black hole
production~\cite{Ahn:2002mj}. The decay of $p$-branes is not
well-understood.  One possibility is that they may decay into lower
dimensional brane-antibrane pairs, leading to a cascade of
branes~\cite{Sen:1999mg}. In any case, there is no reason for them to
decay only to invisible particles, and it is reasonable to expect
their decays, as with BH decays, to be dominated by visible quanta
observable at cosmic ray observatories~\cite{Ahn:2002mj}. With this in
mind, we study the implications of $p$-brane showers for cosmic ray
physics.

Once one entertains the notion of asymmetric compactifications, a wide
variety of possibilities arise, as one can consider the possibility of
several compactification scales.  We consider first the simplest
example in which $n$ flat extra dimensions are divided into two sets,
with $m$ dimensions of length $L$, and $n-m$ larger dimensions of
length $L'$.  Brane production will be significant only in the
presence of Planckian extra dimensions, and so we assume $L \sim
\lstar \equiv \mstar^{-1}$, where $\lstar$ and $\mstar$ are the
fundamental Planck length and mass.  $\mstar$ and the four-dimensional
Planck mass $\mplanck \simeq 1.2 \times 10^{19}~\gev$ are related
by~\cite{Lykken:1999ms}
\begin{equation}
\mplanck^2 = \mstar^{2+n}\, L^{m}\, {L'}^{n-m} \ .
\end{equation}
For simple toroidal compactifications, $L$ and $L'$ are related to
radii by factors of $2\pi$.  Motivated by string/M theory, we will
focus on the cases $n=6,7$.  To facilitate comparison with our earlier
analysis and collider data, we will give results in terms of both
$\mstar$ and $\md = [(2\pi)^n/8\pi]^{1/(n+2)}\,\mstar$.  For $n=6\
(7)$, $\md=2.65 \mstar$ ($2.92 \mstar$).

Scenarios with low values of $n-m$ are already tightly constrained.
Sub-millimeter tests of the gravitational inverse-square law show no
deviation from Newtonian gravity~\cite{Hoyle:2000cv}, yielding
$L'/2\pi \alt 0.2~\text{mm}$.  For $n-m=1\ (2)$ and $L\le \lstar$,
this implies $\mstar\ge 2\times 10^5\ (1.4)~\tev$. Additionally, in
the presence of large extra dimensions, the usual four-dimensional
graviton is complemented by a tower of Kaluza-Klein (KK) states,
corresponding to the new available phase space in the bulk. The
requirement that the neutrino signal of supernova 1987A not be unduly
shortened by the emission of KK modes into the part of the bulk with
large extra dimensions also bounds the compactification
scale~\cite{Cullen:1999hc}.  Such limits are further strengthened by
constraints on KK graviton decay in typical astrophysical
environments, yielding $\mstar\gg 10~\tev$ for $n-m\le
3$~\cite{Hall:1999mk}. For $n-m \ge 4$, bounds from colliders imply
$\mstar \agt 300~\gev$~\cite{Anchordoqui:2001cg}.  All of these bounds
are for flat compactifications. Warped compactifications, in which
bounds for small $n-m$ are much less restrictive, will be discussed
below.

We now consider an uncharged, non-rotating $p$-brane with mass $\mp$
that lives in this (4+$n$)-dimensional spacetime and wraps $r$
Planckian dimensions and $p-r$ large extra dimensions.  Such a
$p$-brane is described by the metric~\cite{Gregory:1995qh}
\begin{equation}
ds^2=R^{{\Delta/(p+1)}}\,(-dt^2+dz_i^2)+R^{{(2-q-\Delta)/
(q-1)}}\,dr^2+r^2\,R^{{(1-\Delta)/(q-1)}}\,d\Omega^2_{q}\ ,
\label{1}
\end{equation}
where $z_i$ ($i=1,\dots,p$) are the brane coordinates, $d\Omega^2_{q}$
($q=2+n-p$) is the metric of the $q$-dimensional unit sphere,
$\Delta=[q (p + 1)/(p+q)]^{1/2}$, and $R=1-(r_p/r)^{q-1}$. The
radius $r_p$ is given by
\begin{equation}
r_p = \frac{1}{\sqrt{\pi} \mstar}\, \gamma(n,p)\,
\left({M_{p}\over \mstar V_p}\right)^{\frac{1}{1+n-p}}\ , \label{rp}
\end{equation}
where $V_p = (L/\lstar)^r (L'/\lstar)^{p-r}$ is the volume wrapped by
the $p$-brane in fundamental Planck units, and
\begin{equation}
\gamma(n,p) = \left[ 8\,\Gamma\left( \frac{3+n-p}{2} \right)
\sqrt{\frac{1+p}{(2+n)(2+n-p)}} \ \right] ^ \frac{1}{1+n-p} \ .
\end{equation}

For $p=0$, \eqref{1} reduces to the metric of a (4+$n$)-dimensional
black hole and $r_p$ becomes the Schwarzschild radius~\cite{Myers:un}.
For $p \ge 1$, \eqref{1} has a naked singularity at $r_p$.  Following
Ref.~\cite{Ahn:2002mj}, we assume that this curvature singularity is
smoothed out by the core of the $p$-brane. We also assume that a BH or
$p$-brane is formed when two partons $i$, $j$ with center-of-mass
energy $\sqrt{\hat{s}}$ scatter with impact parameter $b \le r_p$,
leading to the geometric cross section
\begin{equation}
\hat{\sigma}_{ij\to p\text{-brane}} (\sqrt{\hat{s}}) = \pi r_p^2 \ ,
\end{equation}
where $r_p$ is given by \eqref{rp} with $M_p = \sqrt{\hat{s}}$.

Of interest for this work is the parameter space for which a $p$-brane
cross section dominates the BH cross section. The ratio of these
is~\cite{Ahn:2002mj}
\begin{equation}
\label{ratio}
\Sigma(\hat{s};n,m,p,r)\equiv{\hat{\sigma}_{ij\to p\text{-brane}}
\over \hat{\sigma}_{ij\to BH}} =
\left(\frac{\mplanck}{\mstar}\right)^{-\alpha}
\left({L\over \lstar}\right)^{-\beta}{\gamma(n,p)^2\over\gamma(n,0)^2}
\left({\hat{s}\over M^2_*}\right)^{{p \over (1+n) (1+n-p)}}\,,
\end{equation}
where
\begin{equation}
\alpha={4(p-r)\over (n-m)(1+n-p)}\ge 0\ ,\qquad
\beta={2(nr-mp)\over (n-m)(1+n-p)}\ge 0\ .
\end{equation}
For TeV-scale gravity with $\mstar \ll \mplanck$, $p$-brane production
is negligible relative to BH production unless $p=r$, \ie, the
$p$-brane wraps only Planck size dimensions. (It is also suppressed
for symmetric compactifications with $L=L'$.) In this case,
\eqref{ratio} simplifies to
\begin{equation}
\Sigma(\hat{s};n,m,p,p) = \left({L\over
\lstar}\right)^{-{2p\over 1+n-p}}{\gamma(n,p)^2\over\gamma(n,0)^2}
\left({\hat{s}\over \mstar^2}\right)^{{p\over (1+n) (1+n-p)}}\ .
\label{sig2}
\end{equation}

As can be seen from \eqref{sig2}, and as noted in~\cite{Ahn:2002mj},
$p$-brane production significantly enhances BH production only if $L
\alt \lstar$.  The enhancement results from wrapping on small
dimensions and is a consequence of the dependence of $r_p$ solely on
the density of the $p$-brane; thus, for a given mass, the density and
radius $r_p$ increase with decreasing $L$~\cite{ackn}. On the other
hand, $L$ cannot be much smaller than $\lstar$: in the string-based
low energy Lagrangian, the gauge coupling squared is inversely
proportional to the compactification volume.  A small volume
corresponds to strong coupling and introduces low mass winding
modes. In certain explicit models, these small volumes can be removed
from the gauge sector via a $T$-duality
transformation~\cite{Shiu:1998pa}.  Below, we avoid reference to
specific models, and present results for the generous range $0.1 <
L/\lstar < 10$.

\begin{figure}[tbp]
\begin{minipage}[t]{0.49\textwidth}
\postscript{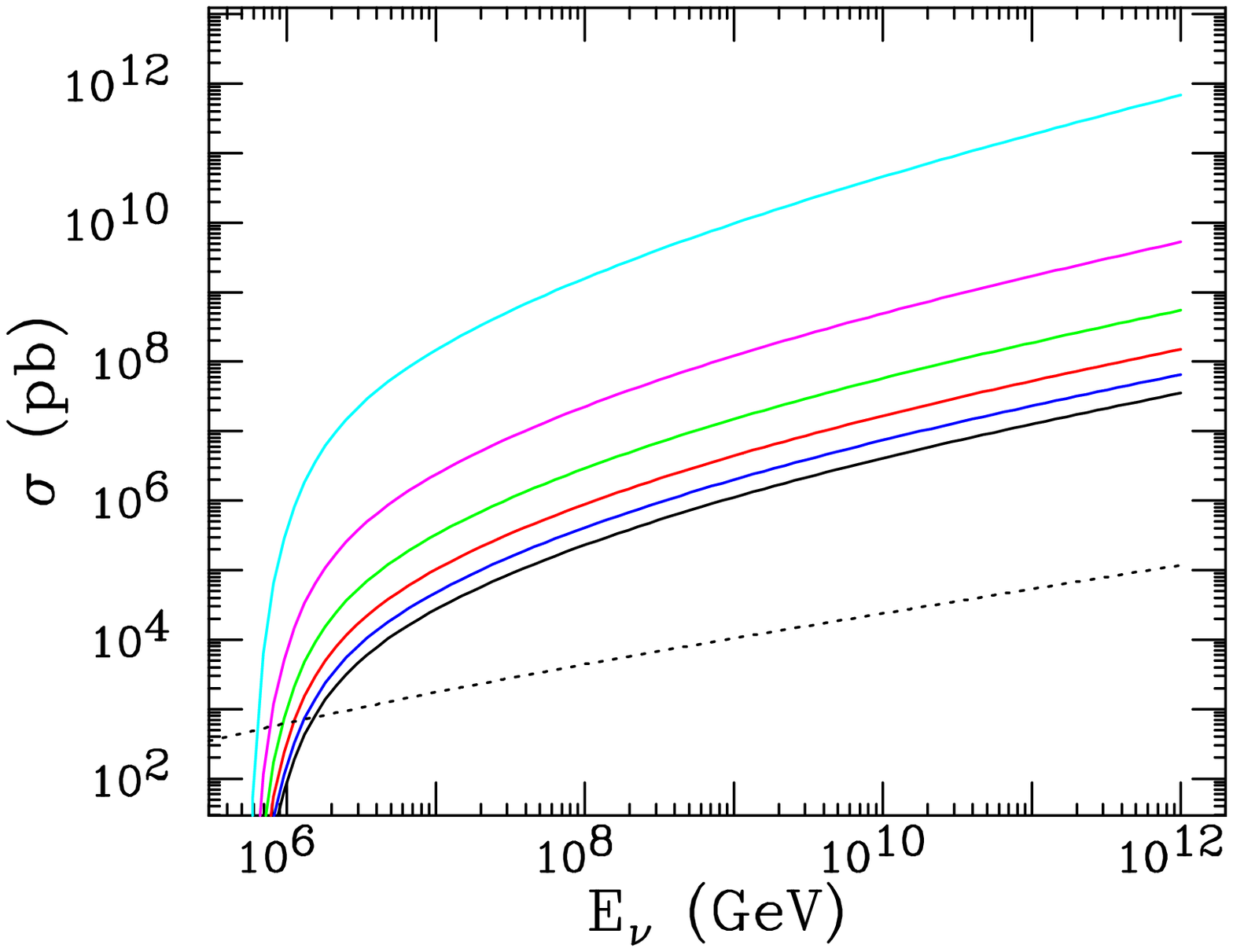}{0.99}
\end{minipage}
\hfill
\begin{minipage}[t]{0.49\textwidth}
\postscript{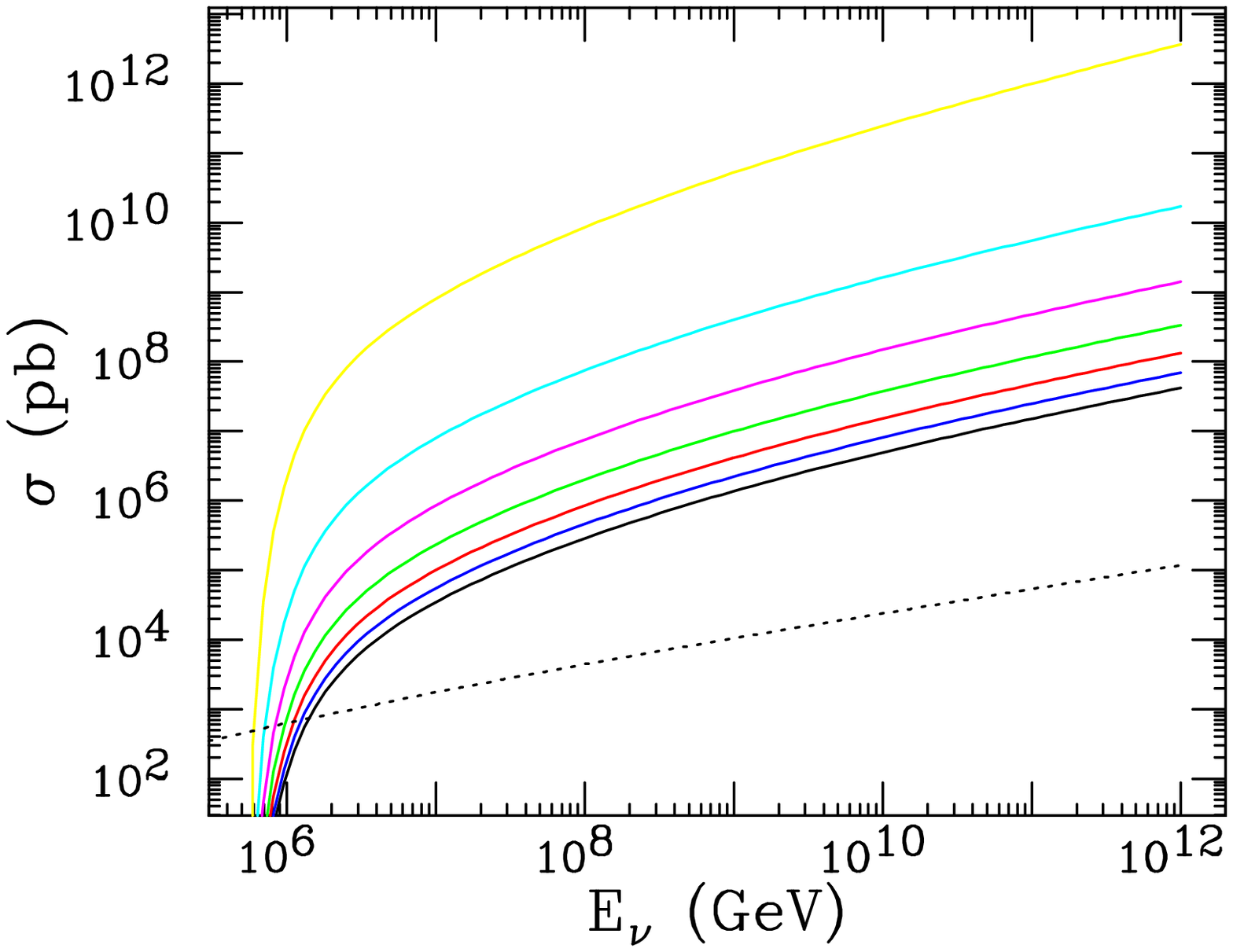}{0.99}
\end{minipage}
\caption{Total cross section $\sigma (\nu N \to \text{brane})$ for
$n=6$ (left), $n=7$ (right), $L/\lstar = 0.25$, $\md= \mpmin =1~\tev$,
and $m = 0, \ldots, n-1$ from below. The standard model cross section
$\sigma(\nu N \to \ell X)$ (dotted) is also shown.
\label{fig:sigma} }
\end{figure}

The cross section for $p$-brane production from neutrino-nucleon
scattering is
\begin{equation}
\label{partonsigma}
\sigma ( \nu N \to p\text{-brane}) =
\sum_i \int_{\mpmin{}^2/s}^1 dx\,
\hat{\sigma}_i ( \sqrt{xs} ) \, f_i (x, Q) \ ,
\end{equation}
where $s = 2 m_N E_{\nu}$, the sum is over all partons in the nucleon,
and the $f_i$ are parton distribution functions.  As
in~\cite{Feng:2001ib,Anchordoqui:2001cg}, we set the momentum transfer
$Q = \min \{ M_p, 10~\tev \}$, where the upper limit is from the
CTEQ5M1 distribution functions~\cite{Lai:2000wy}. Finally, $\mpmin$ is
the minimum $p$-brane mass required for production, which we assume
equal to $\md$.

To obtain the total cross section for brane production, we assume that
$p$-brane production is possible for all $p$, and so the total cross
section is
\begin{equation}
\sigma (\nu N \to \text{brane})  =
 \sum_{p=0}^m \sigma ( \nu N \to p\text{-brane}) \ .
\end{equation}
Total cross sections for brane production by cosmic neutrinos are
given in Fig.~\ref{fig:sigma} for $L/\lstar = 0.25$ and $\md =
1~\tev$.  The lowest solid curves for $m=0$ are for BH production
only, and are greatly enhanced relative to the standard model.  We
see, however, that for small values of $L/\lstar$, even larger cross
sections are possible for $p$-branes, especially for low $n-m$.

It has recently been proposed~\cite{Jain:2002kf} that ultra-high
energy neutrinos interacting via $p$-brane production may provide a
solution to the puzzle of the observed cosmic rays with energies above
$10^{11}~\gev$, \ie, above the Greisen-Zatsepin-Kuzmin (GZK)
limit~\cite{Greisen:1966jv}.  These cosmic ray showers begin high in
the atmosphere, and so require $\nu N$ cross sections of order 100 mb.
We see from Fig.~\ref{fig:sigma}, however, that such cross sections
are approached only for one or two large extra dimensions $(n-m=1,2)$
and $\md \approx 1~\tev$, a region of parameter space excluded by the
sub-mm gravity experiments and astrophysical constraints discussed
above. {\em Cosmic neutrinos with interaction strengths enhanced by
$p$-brane production cannot resolve the GZK paradox in flat
compactification scenarios}.  We will return to the possibility of
warped compactifications below.

While $p$-brane cross sections for $n-m \ge 3$ are irrelevant for the
GZK paradox, they may nevertheless enhance deep shower rates, with
strong implications for cosmic ray experiments.  For asymmetric
spacetimes with $n-m \ge 3$, the event rate for deep showers is
\begin{equation}
{\cal N} = \int\, dE_\nu\, N_A\, \frac{d\Phi}{dE_\nu}\,
\sigma ( \nu N \to \text{brane}) \, A(E_\nu)\, T \ , \label{events}
\end{equation}
where $N_A = 6.022 \times 10^{23}$ is Avogadro's number,
$d\Phi/dE_\nu$ is the neutrino flux, $A(E_\nu)$ is the acceptance for
quasi-horizontal showers in cm$^3$ water equivalent steradians, and
$T$ is the experiment's running time.  For the neutrino flux, we
consider the conservative cosmogenic flux produced by interactions of
the observed ultra-high energy protons with the cosmic microwave
background.  Specifically, as in our previous
paper~\cite{Anchordoqui:2001cg}, we adopt the estimates of Protheroe
and Johnson with an injection spectrum with cutoff energy $3\times
10^{12}~\gev$~\cite{Protheroe:1996ft}.  Additional fluxes are possible
and would only strengthen the conclusions below.

The AGASA Collaboration has searched for deeply penetrating
showers~\cite{agasa}.  An estimate of the AGASA acceptance for deeply
penetrating events is given in~\cite{Anchordoqui:2001cg}. In
$T=1710.5$ days of data taking, they find 1 event with an expected
background of 1.72, leading to a 95\% CL limit of 3.5 events from
$p$-brane creation.

\begin{figure}[tbp]
\begin{minipage}[t]{0.49\textwidth}
\postscript{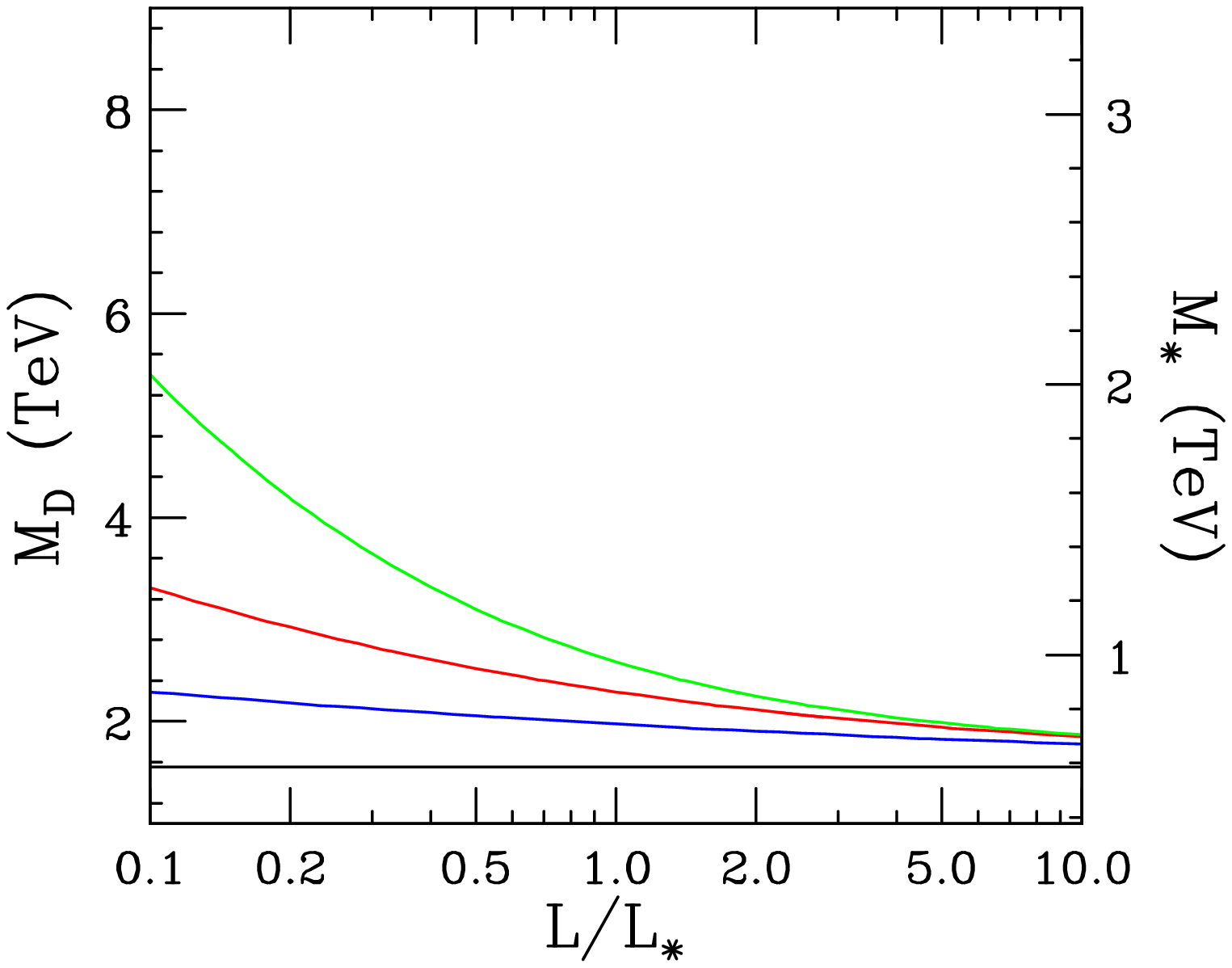}{0.99}
\end{minipage}
\hfill
\begin{minipage}[t]{0.49\textwidth}
\postscript{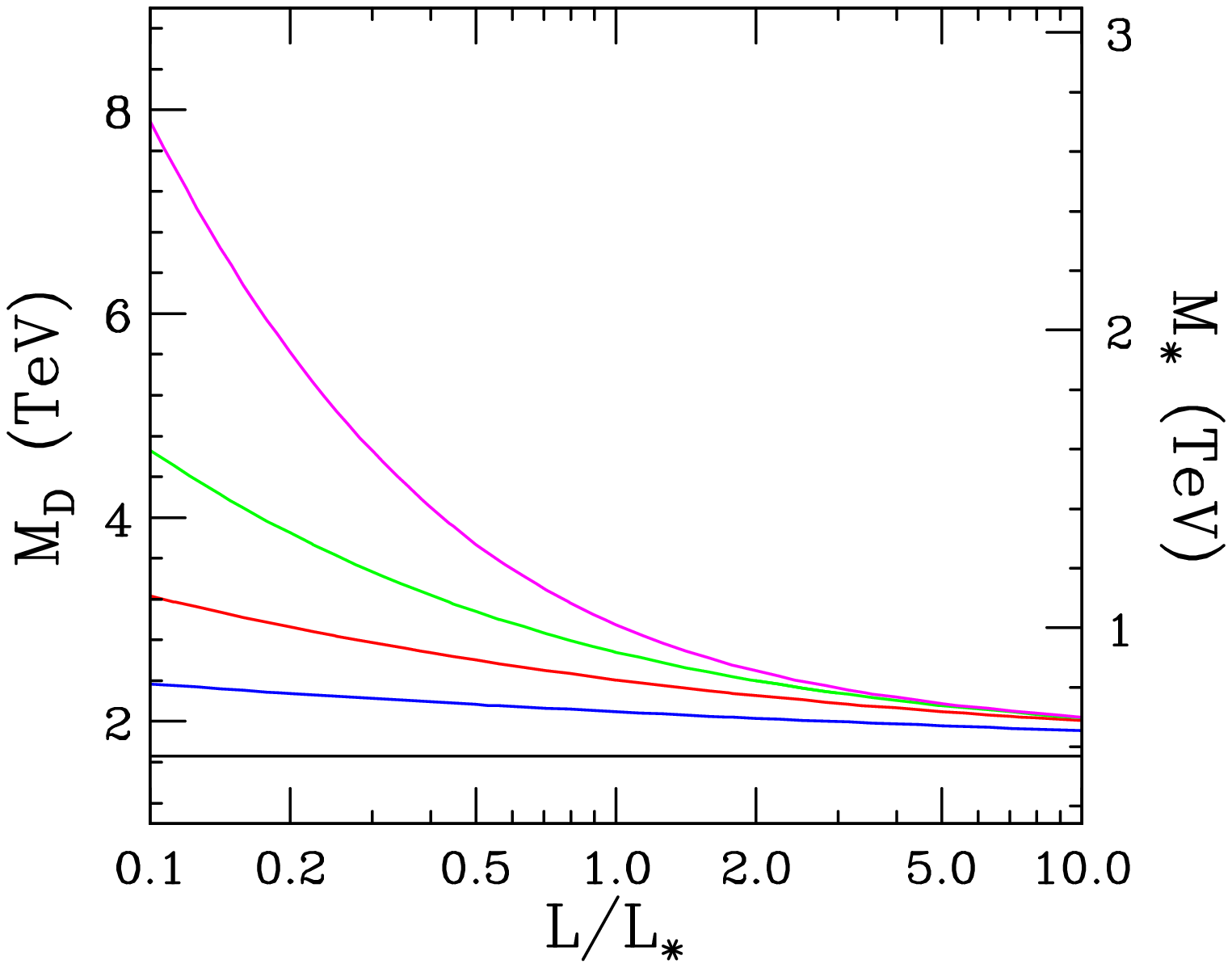}{0.99}
\end{minipage}
\caption{95\% CL lower bounds on $\md$ in asymmetric compactification
scenarios from the absence of $p$-brane-induced deep showers at AGASA.
Bounds are given for various Planckian compactification lengths $L$,
$n=6$ (left) and $n=7$ (right), $m=0,\ldots, n-3$ from below, and
$\mpmin = \md$. For each $m$, contributions from $p=0,\ldots, m$ are
summed.
\label{fig:limits} }
\end{figure}

The absence of evidence for deeply penetrating showers then places
bounds on the parameter space of asymmetric compactifications.  These
bounds are given in Fig.~\ref{fig:limits}, and the results can be
summarized as follows: (1) For $m=0$, only 0-branes (BHs) are
produced.  The bounds on $\md$ are therefore independent of
$L/\lstar$, and we recover the constraint $\md \agt 1.6~\tev$, first
given in~\cite{Anchordoqui:2001cg}. (2) For $L/\lstar\to \infty$, the
contribution from $p\ne 0$ vanishes, and all limits asymptotically
approach the BH bound. (3) Even for $L\simeq \lstar$, the limits on
$\md$ from $p$-brane production are significantly enhanced above
limits from BH production alone, and are as large as $3~\tev$. (4) For
smaller values of $L/\lstar$, the lower bounds on $\md$ rise
dramatically. (5) The Auger Observatory, scheduled for completion by
2004 with an acceptance roughly 30-100 times that of AGASA, will
provide an extremely sensitive probe of asymmetric compactifications.

We now return to scenarios with $n-m = 1$ or 2, but consider the
possibility that these dimensions are warped~\cite{Giddings:2001yu}.
Although no explicit models are available, these scenarios may evade
the stringent constraints on $\md$ from astrophysics and Newtonian
gravity.  At the same time, the cross sections for $p$-brane
production may be as in the flat compactification case if the
curvature length scales are large compared with $r_p$.  In these
scenarios, can $p$-brane production provide an explanation for cosmic
rays above the GZK cutoff?

This possibility is constrained by at least three considerations.
First, as in the $n-m \ge 3$ cases, these scenarios are limited by the
absence of deeply penetrating showers at AGASA.  The expected deep
shower event rate is determined essentially as before, but now cross
sections may be so large that showers begin high in the atmosphere and
so do not contribute to deep shower rates.  The atmospheric depth for
quasi-horizontal showers with zenith angle $70^{\circ}$ is about
$3000~\g/\cm^2$.  This interaction length corresponds to a cross
section of $\sigma_{\nu N}= 0.56~\mb$.  We determine deep shower rates
assuming conservatively that only neutrinos with total cross sections
below 0.56 mb contribute~\cite{Tyler:2000gt}.

Second, if $\sigma (\nu N) \agt 100~\mb$ for $E_\nu> 10^{11}~\gev$,
then one expects $\sigma(\nu N) \sim 1$ to 10 mb for $E_\nu \sim
10^9~\gev$. (See Fig.~\ref{fig:sigma}.) This implies that cosmic
neutrinos should produce $p$-brane showers (akin to black hole
showers~\cite{Anchordoqui:2001ei}), but with primaries with mean free
paths of $\lambda_{\nu\text{-air}}\sim 4-30$ times larger than
$\lambda_{p\text{-air}}$. Such {\em moderately} penetrating showers
were discussed in~\cite{Anchordoqui:2000uh}. Because this cross
section would occur near the peak of the cosmogenic
flux~\cite{Protheroe:1996ft}, such showers will be copiously produced
and should be observed at cosmic ray detectors. This is an important
test for these scenarios --- an abundance of such moderately
penetrating showers have not been reported to date.

Third, very large cross sections lead, via a dispersion relation, to
large deviations in standard model predictions at lower
energies~\cite{Goldberg:1998pv}. With the cross sections of
Fig.~\ref{fig:sigma}, it is straightforward to apply the results
of~\cite{Goldberg:1998pv} to show that cross sections
\begin{equation}
\sigma_{\nu N}(10^{11}~\gev) \ge 300~\mb 
\label{gw}
\end{equation}
lead to $\sim 100\%$ corrections to, \eg, neutrino properties at
energies $\sim 100~\gev$.

\begin{figure}[tbp]
\begin{minipage}[t]{0.49\textwidth}
\postscript{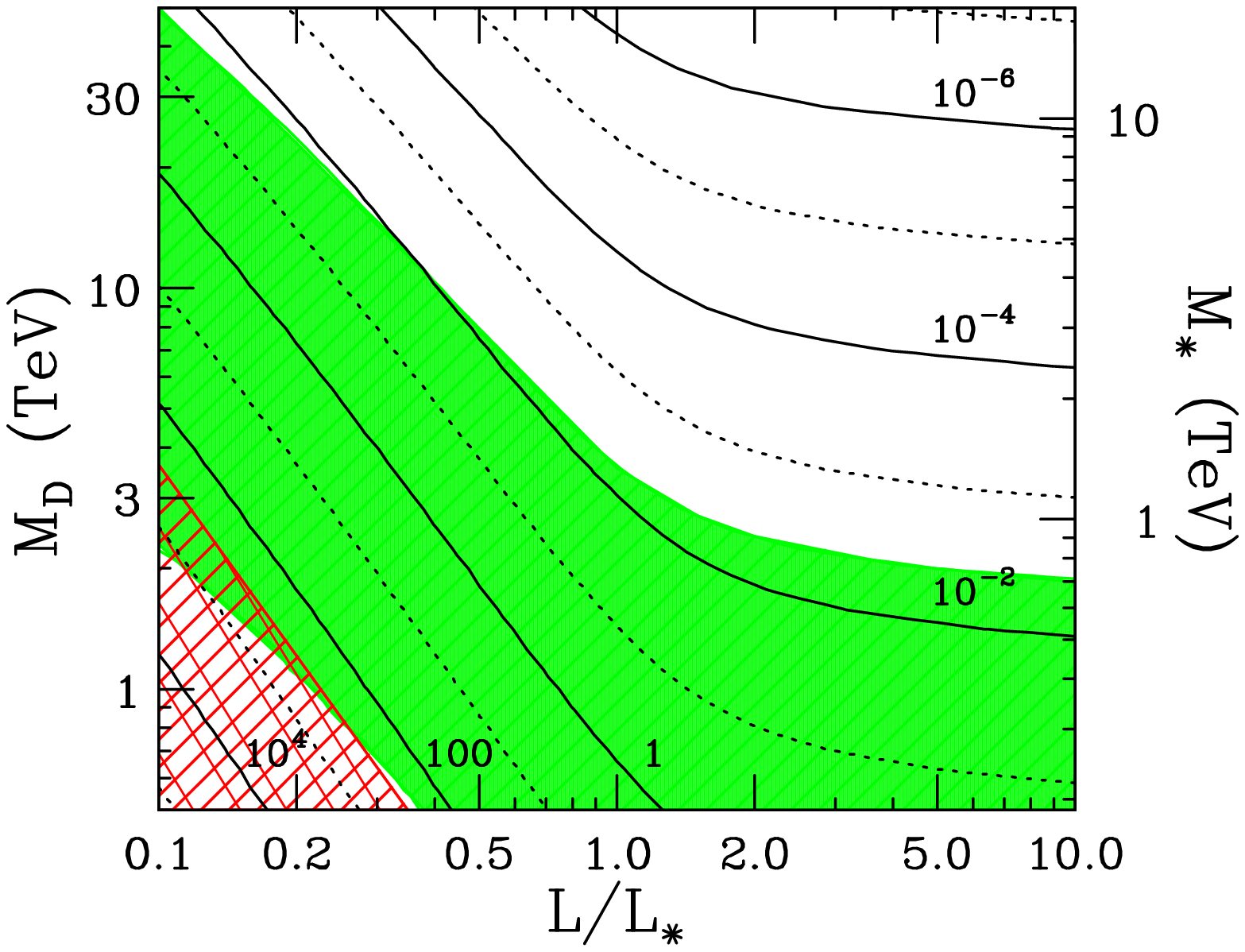}{0.99}
\end{minipage}
\hfill
\begin{minipage}[t]{0.49\textwidth}
\postscript{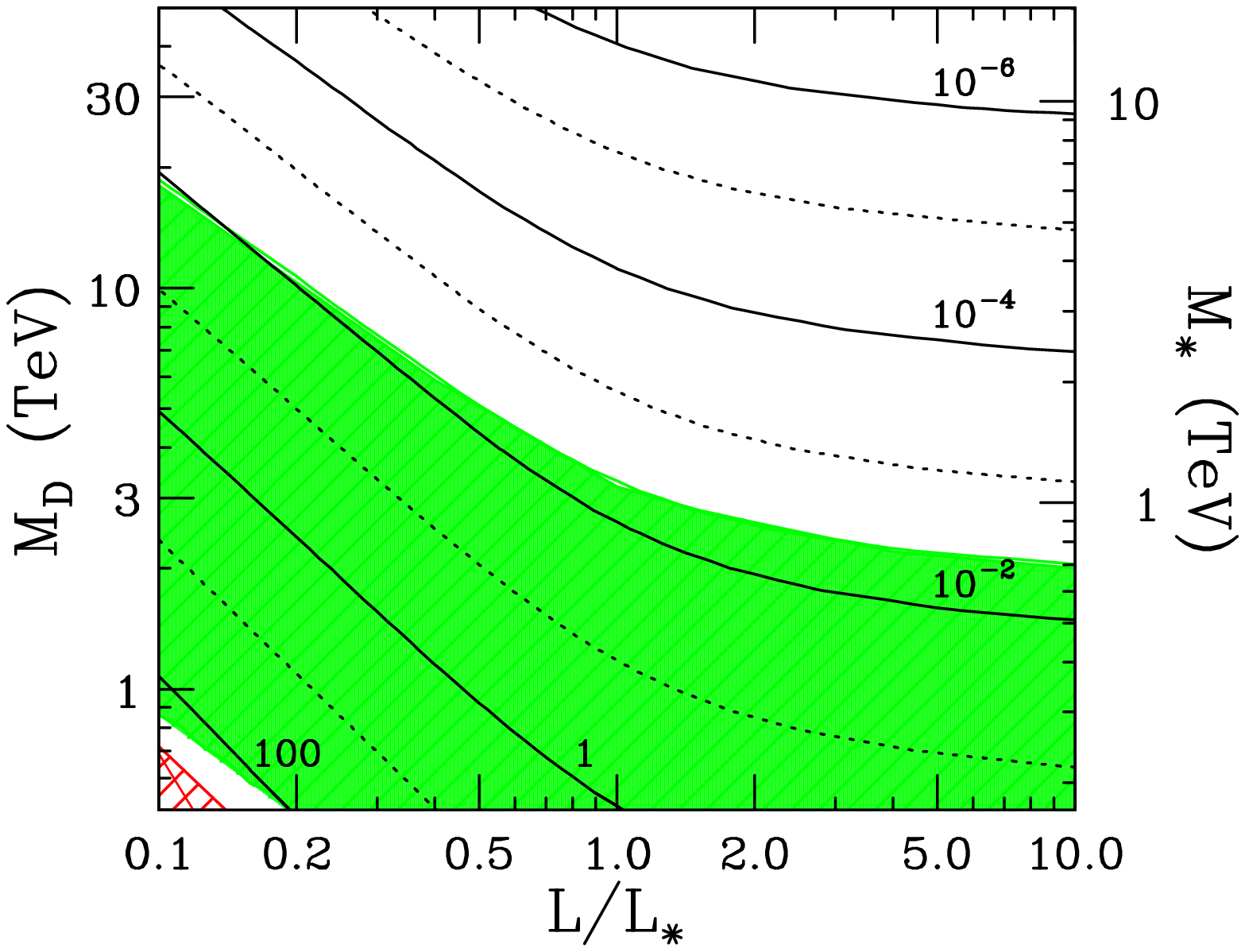}{0.99}
\end{minipage}
\caption{Contours of total cross section $\sigma ( \nu N \to
\text{brane})$ (in mb) at $E_{\nu} = 10^{11}~\gev$ in the $(L/\lstar,
\md)$ plane for $(n,m) = (6,5)$ (left) and $(7,5)$ (right).  The
shaded region is excluded by the non-observation of deeply penetrating
showers at AGASA. In the hatched region, large corrections to standard
model physics at 100 GeV energies are expected (see text).
\label{fig:gzk} }
\end{figure}

Cross sections $\sigma (\nu N)$ at $E_\nu = 10^{11}~\gev$ are given in
Fig.~\ref{fig:gzk} for two scenarios with $n-m = 1,2$.  [Results for
$(n,m) = (7,6)$ are very similar to those for $(6,5)$.]  The shaded
area is excluded by the AGASA bound on deeply penetrating showers. For
large $\md$, cross sections are sufficiently suppressed to eliminate
large deep shower rates.  The upper boundary of this shaded region
agrees with existing limits~\cite{Tyler:2000gt}.  The AGASA constraint
may also be evaded in the lower left corners, where cross sections are
so large that $p$-brane showers develop high in the atmosphere and
appear hadronic. These regions predict moderately penetrating showers.
In addition, cross sections in this region are typically extremely
large, and so require modifications to standard model physics at lower
energies as discussed above.  The region satisfying \eqref{gw} is
cross-hatched in Fig.~\ref{fig:gzk}.

As can be seen in Fig.~\ref{fig:gzk}, in some regions of parameter
space, cross sections of $\sim 100~\mb$ are sufficient to mimic the
highest energy cosmic rays.  However, all of the desired parameter
space with $\sigma (\nu N) < 100~\mb$ is excluded by the
non-observation of deeply penetrating showers at AGASA.  Regions with
$\sigma(\nu N) > 100~\mb$ evade this constraint, but predict
moderately penetrating showers, and large corrections to standard
model physics at $\sim 100~\gev$ energies. A GZK solution also
requires small extra dimensions with size considerably below the
fundamental Planck length, as well as low $\md$ values subject to
collider probes.

In summary, we have considered the implications of $p$-brane
production by ultra-high energy neutrinos. Current AGASA data imply
multi-TeV bounds on $\md$, the strongest bounds on asymmetric
compactifications for $n-m \ge 4$.  Auger, with a projected
sensitivity 30 to 100 times that of AGASA, will either discover
$p$-brane showers or significantly strain attempts to identify the
weak and fundamental Planck scales in these scenarios.  For flat
compactifications, astrophysical and sub-mm gravity constraints
exclude a $p$-brane explanation of super-GZK events.  For warped
compactifications, much of the potential parameter space is excluded
by AGASA data.  The remaining scenarios require low $\md$ and small
extra dimensions $L<0.2\lstar$, leading to strong coupling effects in
the underlying stringy regime.  These solutions also predict $\sim
100\%$ corrections to standard model neutrino physics at the 100 GeV
scale, and moderately penetrating showers, not reported to date.
These considerations leave very little room for explaining super-GZK
events with $p$-brane physics.

\begin{acknowledgments}
We thank Eun-Joo Ahn, Marco Cavagli\`a, Carlos Nu\~nez, Angela Olinto,
Al Shapere, and Tom Taylor for informative discussions.  The work of
LAA and HG has been partially supported by the US National Science
Foundation under grants No.\ PHY--9972170 and No.\ PHY--0073034,
respectively. The work of JLF was supported in part by the Department
of Energy under cooperative research agreement DF--FC02--94ER40818.
\end{acknowledgments}


\end{document}